# Comment on the Theory of the Stretching Experiments of RNA in Water

E. G. D. Cohen
The Rockefeller University
New York NY 10065

**Abstract**
It is argued that the stretching experiments done on RNA in water can be described as a reversible process by Classical Thermodynamics.

Chemical reactions in water have been treated in textbooks of Classical Thermodynamics as taking place reversibly, in spite of the possible irreversible nature of such reactions.

In the case of the stretching of an RNA molecule in water, the RNA molecule is not an autonomous system, but a very complicated activated complex consisting of an RNA molecule and the surrounding water. During its stretching, hydrogen bonds are broken, so that water molecules invade the RNA molecule to replace the broken bonds. This means that the heat bath penetrates the RNA molecule, thus incorporating the RNA molecule effectively into the heat bath. This does not occur in Classical Thermodynamics, since there the heat baths only provides heat not molecules.

Furthermore, the complicated hysteresis loops observed in the experiments, are relatively small compared to $k_B T$ -- where T is the temperature of the heat bath and $k_B$ Boltzmann's constant -- so that they merely reflect the complexity of the super-system of an RNA molecule in water.

In two previous papers [1,2] it was suggested that the agreement between the experiments of Bustamante et al.-- as well as of many others -- and the Jarzynski Equality was due to the fact that the Jarzynski Equality is actually only valid for a reversible process, as is shown in [1].

This can be seen as follows. The Jarzynski Equality reads $<e^{-\beta W}>_{irr} = e^{-\beta \Delta F}$. Here W on the left hand side of this equation, is the irreversible mechanical work $W_{irr}$ done on the RNA system in water, when the external stretching forces bring this system from one state to another, averaged over all *ir*reversible paths with weights $e^{-\beta W_{irr}}$. Here $\beta = 1/k_B T$, where $k_B$ is Boltzmann's constant and T the absolute temperature of the system as well as that of the heat bath coupled to it. $\Delta F$ is the difference of the Helmholtz free energy of the RNA complex, due to the work done on it. However, the average of the work is taken over all possible *ir*reversible work processes. Then, in Classical Thermodynamics, by the definition of an irreversible process -- where work is done on a system that does *not* stay in thermal equilibrium at the heat bath's temperature T during the entire work process -- only *in*equalities can be obtained, as e.g. in the Carnot-Clausius relation. Thus, Jarzynski completely ignores that for the irreversible processes, which he considers, only *in*equalities exist in Classical Thermodynamics.

The two papers [1,2] mentioned above did not discuss Crooks' theorem. However, since this theorem also contains Jarzynski's equality the same objection applies as for Jarzynski's equality.

Therefore, for the explanation of the RNA in water experiments, neither Jarzynski's equality nor Crooks' theorem are needed, since they are both also valid for reversible processes, so that then a Boltzmann factor, with the heat bath temperature T can be used. This may explain the agreement of these two incorrect equalities with the experiments. For more details I refer to the two mention papers cited above.

Summarizing: Since the RNA in water experiments are effectively done on the very complex system of RNA and water, they can be understood on the basis Classical Reversible Thermodynamics.